\documentclass[a4paper,conference,final]{IEEEtran}
\usepackage{siunitx}
\DeclareSIUnit \dBm {dBm}
\DeclareSIUnit \dB {dB} 
\DeclareSIUnit \dBi {dBi} 
\DeclareSIUnit \Kbps {Kbps}
\DeclareSIUnit \Mbps {Mbps}
\DeclareSIUnit \Gbps {Gbps}
\DeclareSIUnit \kBps {kBps}
\DeclareSIUnit \MBps {MBps}
\DeclareSIUnit \GBps {GBps}

\usepackage{cite}      
\usepackage[caption=false, font=footnotesize]{subfig}
\usepackage[nolist,nohyperlinks]{acronym}
\usepackage[pdftex]{graphicx}
\usepackage{url}       
\usepackage{amsfonts}
\usepackage{amsmath}
\usepackage{times}
\usepackage{algpseudocode}
\usepackage{fourier}
\usepackage{algorithm}
\usepackage{enumerate}
\usepackage{siunitx}
\usepackage{multirow}
\usepackage{tikz}
\usepackage{array}
\usepackage{gensymb}
\usepackage{physics}
\DeclareMathOperator*{\argmax}{argmax}
\newcommand*{\argmaxl}{\argmax\limits}
\newcolumntype{P}[1]{>{\centering\arraybackslash}p{#1}}
\newcolumntype{M}[1]{>{\centering\arraybackslash}m{#1}}

\begin{document}
\title{Beam Alignment for Millimetre Wave Links with Motion Prediction of Autonomous Vehicles}
\author{\IEEEauthorblockN{Ioannis Mavromatis, Andrea Tassi, Robert J. Piechocki, and Andrew Nix}
  \IEEEauthorblockA{Department of Electrical and Electronic Engineering, University of Bristol, UK \\ Emails: \{ioan.mavromatis, a.tassi, r.j.piechocki, andy.nix\}@bristol.ac.uk}
}

\maketitle
\begin{abstract}
Intelligent Transportation Systems (ITSs) require ultra-low end-to-end delays and multi-gigabit-per-second data transmission. Millimetre Waves (mmWaves) communications can fulfil these requirements. However, the increased mobility of Connected and Autonomous Vehicles (CAVs), requires frequent beamforming - thus introducing increased overhead. In this paper, a new beamforming algorithm is proposed able to achieve overhead-free beamforming training. Leveraging from the CAVs sensory data, broadcast with Dedicated Short Range Communications (DSRC) beacons, the position and the motion of a CAV can be estimated and beamform accordingly. To minimise the position errors, an analysis of the distinct error components was presented. The network performance is further enhanced by adapting the antenna beamwidth with respect to the position error. Our algorithm outperforms the legacy IEEE 802.11ad approach proving it a viable solution for the future ITS applications and services. 
\end{abstract}

\begin{IEEEkeywords}
Connected Autonomous Vehicles, mmWave, Beamwidth Optimisation, Beamforming, Heterogeneity, MAC layer, Vehicle-to-Infrastructure.
\end{IEEEkeywords}

\section{Introduction}
Connected and Autonomous Vehicles (CAVs), when fully commercialised, will provide significant user convenience and safety as well as reduced pollution and fuel consumption. Self-driving vehicles will be members of the ecosystem of \emph{Next-Generation Intelligent Transportation Systems (ITSs)}. This ecosystem will provide services such as transit management, emergency management, multi-modal commuting, etc.~\cite{its_appl}. Massive amount of data will be generated and exchanged between the various entities of the system. As an example, it is expected that just one self-driving vehicle will generate more than a gigabit-per-second of sensory data~\cite{gigabit_second}.

The strict Quality-of-Service (QoS) constraints of the next-generation automotive applications such as 'tactile-like interactions' and gigabit-per-second throughput~\cite{qos_req} can be effectively supported using \emph{Millimetre Wave (mmWave) communications}. However, their propagation characteristics combined with the increased vehicle mobility lead to performance degradation (e.g. due to Doppler shifts because of misalignments~\cite{80211ad_anal}). To that extent, mmWave adoption for vehicular communications require smarter ways to tackle the problems arose. CAVs, as smart entities within an ITS are equipped with numerous sensors and Radio Access Technology (RAT) interfaces that combined can potentially provide a solution to the aforementioned problems. 

Based on this idea we propose a smart, network-based mmWave beamforming strategy for Vehicle-to-Infrastructure (V2I) links, able to reduce beam misalignments under the highly dynamic vehicular networks. More specifically, we will focus on the frequency band of \SI{60}{\giga\hertz} and how the position and direction information broadcast can be utilised to improve the performance of a mmWave system by enhancing its beamforming training algorithm.

Referring to IEEE 802.11ad~\cite{standard}, the beamforming process requires a bidirectional frame transmission. According to~\cite{beamforming_delay} the beam switching for a phased-array antenna is almost instantaneous ($\simeq$\SI{50}{\nano\second}). Therefore, the beamforming delay is entirely related with the number of frames exchanged. To that extent, authors in~\cite{exhaustive} introduced a new codebook design scheme able to accelerate the beam training process, reduce the overhead and improve the system performance. However the beam training is still performed in quasi-omnidirectional mode. Therefore, the system cannot compensate with the increased mobility due to the Doppler Spread.

For mobile environments, more frequent beamforming is required, increasing the training overhead. A viable solution is a network operating in a heterogeneous manner, exchanging the mmWave training information out-of-band. Dedicated Short Range Communications (DSRC) can achieve high packet delivery ratio for short distances ($\le$\SI{200}{\meter}), even under urban environments~\cite{packet_delivery}. To that extent, authors in~\cite{beam_design} designed a heterogeneous network consisting of DSRC and mmWave RATs. Facilitating the position information sent within a DSRC beacon they predict the movement of the vehicle and perform beamforming. However, positioning errors were not taken into account, vehicle speed was constant and no complex manoeuvres were considered. On a similar approach, our algorithm can enhance performance by fusing the position, motion and velocity data from a vehicle.

The position information of the vehicle is most commonly acquired via Global Positioning System (GPS) and is not perfect. This will lead to imperfections and slight misalignments in our algorithm. What is more, GPS error is not consistent as observed in the 3D space. Easting, northing, and elevation errors might vary and will influence differently the system. In such manner, the different GPS error components will be further analysed deriving equations for thee sensitivity of the system on each individual error. As shown in~\cite{beam_opt}, when misalignment errors introduced there is an optimal non-zero beamwidth that maximises the system performance. In this fashion, and taking into consideration the sensitivity analysis mentioned before, equations for the beamwidth optimisation with respect to the error will be derived.

This paper is organised as follows. In Sec.~\ref{sec:11ad_beamforming}, a detailed explanation of the overhead computation for IEEE 802.11ad will be given as well as the reasons that the legacy beamforming technique is not a viable solution for vehicular communications. The proposed algorithm is described in Sec.~\ref{sec:het_net},  presenting the required models and the algorithmic steps. The section will be concluded with the positioning error analysis and the beamwidth optimisation problem. Simulated and numerical results will be presented and discussed in Section~\ref{sec:results} and the work will be concluded in Section~\ref{sec:conclusions}, with ideas for future research.

\section{Traditional Beamforming with IEEE 802.11ad} \label{sec:11ad_beamforming}
IEEE 802.11ad is the dominant standard for mmWave communications~\cite{standard}. In the standard, the interval between two beacon frames is defined as the \emph{Beacon Interval (BI)}. BI is subdivided in two access periods. The first one, called as \emph{Beacon Header Interval (BHI)} facilitates the exchange of management information and network announcements. The second period is responsible for the data transmission and is called \emph{Data Transmission Interval (DTI)}. BHI is further subdivided in shorter access periods. These are: the \emph{Beacon Transmission Interval (BTI)}, used for network announcement and beamforming training, the \emph{Association Beamforming Tranining (A-BFT)} where antennas are trained and paired with the Personal Basic Service Set (PBSS) Control Point (PCP)/AP, and finally the \emph{Announcement Transmission Interval (ATI)} during which management information is exchanged with the associated stations (Fig.~\ref{fig:bi_frame}).

%The Personal Basic Service Set (PBSS) of IEEE 802.11ad, as described in the standard consists of one PBSS Control Point (PCP) and $N\leq254$ number of Directional Multi-Gigabit (DMG) stations.

The propagation characteristics in the frequency band of \SI{60}{GHz} result to severe signal attenuation during quasi-omnidirectional communications. Therefore, the MAC layer of IEEE 802.11ad introduces the concept of "virtual" antenna sectors. These sectors divide the azimuth plane into a number of sectors, depending on the type of the device used (e.g. a PCP/AP will utilise more "virtual" sectors than a handheld device). These "virtual" sectors can be further subdivided with respect to the minimum beamwidth of the antenna creating a \emph{two-layer model} representation of the beams and the beamforming process (Fig.~\ref{fig:bi_frame}). In such manner, the beamforming in IEEE 802.11ad is performed in two phases exchanging a bidirectional frame sequence. 

The first phase (first-layer), is called  \emph{Sector Level Sweep (SLS)}. SLS trains a transmitting antenna using an iterative sweeping process based on the strongest Signal-to-Noise Ratio (SNR). Later, during \emph{Beam Refinement Phase (BRP)} (second-layer), the receiving antenna is trained and the beams are further refined, choosing finally a pair of beams able to compensate with the channel losses at \SI{60}{GHz}.

\subsection{Overhead analysis of IEEE 802.11ad beamforming process}\label{sub:overhead}

Consider a typical vehicular network with one PCP/AP on the side of the road and a number of mobile devices travelling on this road. As it was mentioned before, most of the waisted time comes from the training frames exchanged. 

\begin{figure*}[t]     
\centering
\includegraphics[width=1\textwidth]{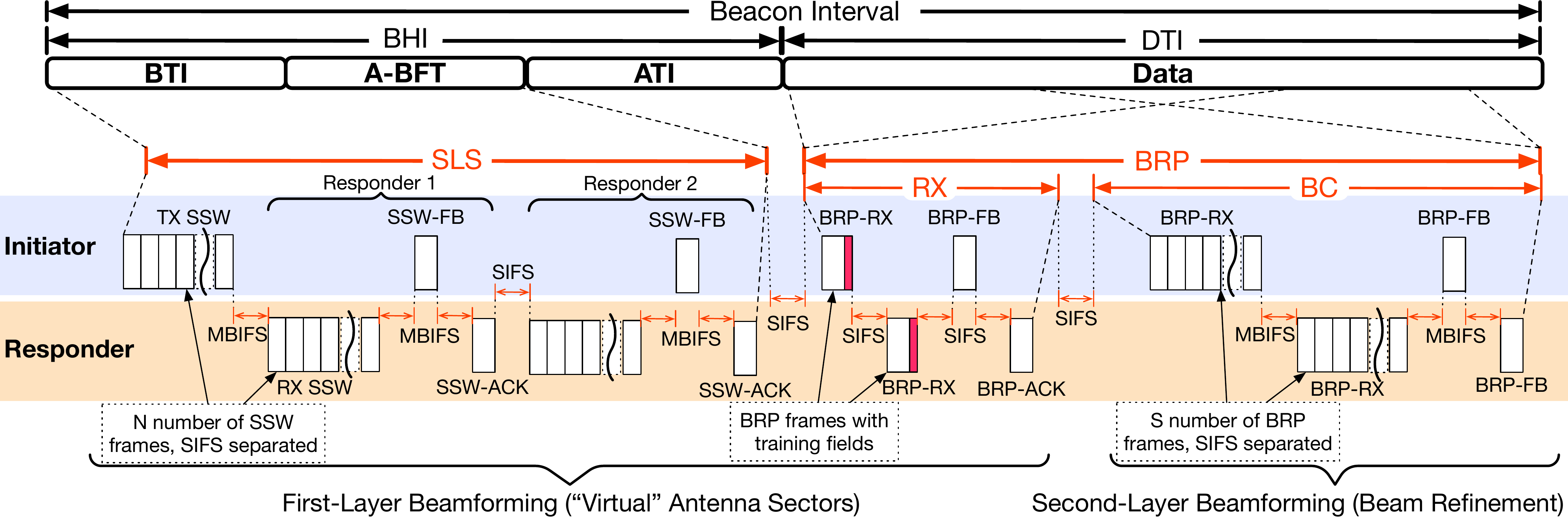}
    \caption{The Beacon Interval and the \emph{two-layer} beamforming: An example of two stations participating in the beamforming process. The frames exchanged, the interframe spacing and the intervals that each phase requires.}
    \label{fig:bi_frame}
\end{figure*}

The SLS is executed during BTI and A-BFT in four steps (Fig.~\ref{fig:bi_frame}): 1) The \emph{initiator} transmits one directional training frame per sector while the receiving devices (\emph{responders}) listen in quasi-omnidirectional mode. 2) The responders reply with a directional frame throughout all their sectors. 3) Feedback information is transmitted from the initiator with the Sector Sweep Feedback (SSW-FB) frames and 4) is acknowledged by the responders with the Sector Sweep Acknowledges (SSW-ACKs). When more than one stations exist in the coverage region of PCP/AP, the slotted A-BFT introduces a contention-based response period with values between $U(4,8)$ based on a uniform random distribution~\cite{standard}. Allocating one slot per station, more than one devices can respond to the beacon sweep reducing the collisions of RX-SSW frames. The total time required for a SLS can be calculated as:
\begin{equation}
T_{SLS} = K \cdot T_{TX-SSW} + N \cdot (K\cdot T_{RX-SSW} + T_{SSW-FB/ACK}) + T_{IFS}
\end{equation}
where $T_{TX-SSW}$, $T_{RX-SSW}$, $T_{SSW-FB}$ and $T_{SSW-ACK}$ are the required time for the different frames exchanged, $K$ is the number of "virtual" sectors (in this case, same number was assumed for all the devices), $N$ is the number of stations around the PCP/AP and $T_{IFS}$ is the total interframe spacing time and is equal to $T_{IFS}= (N+1)(K-1) \cdot T_{SBIFS} + 3 \cdot T_{MBIFS}$.

During the BRP phase, multiple configurations can be tested with one frame transmission, reducing the overhead compared to SLS. BRP can be divided in two subphases. At first, the best RX antenna sector is found by exchanging BRP frames appending \emph{transmit and receive training fields (TRN-TX/RX)} and followed by a feedback frame and an acknowledgement. The time required is given as follows:
\begin{equation}
T_{RX} = 2 \cdot T_{BRP} + T_{BRP-FB} + T_{BRP-ACK} + 3 \cdot T_{SIFS}
\end{equation}
By the end of the above process, a pair of "virtual" antenna sectors is chosen with respect to the higher SNR. During the second subphase (\emph{Beam Combining (BC)}), the beams are further refined. A set of pairwise antenna weight vector combinations is tested between the two devices as a directional link between the devices is already established. The time required can be calculated as:
\begin{equation}
T_{BC} = S \cdot T_{BRP} + T_{BRP-FB/ACK} + 3 \cdot T_{MBIFS} + (S-1) \cdot T_{SIFS}
\end{equation}
where $S$ is the number of antenna weight vector combinations tested. Finally, the total time required for the beamforming process is as follows:
\begin{equation}
T_{all} = T_{SLS} + T_{RX} + T_{BC}+2 \cdot T_{SIFS}
\end{equation}

This section introduced an approximation of the time required for the beamforming process of $N$ number of devices. However, a perfect channel with zero frame loss was considered and BRP can be more complicated (e.g. a BRP setup subphase might be required if BRP does not follow an SSW-ACK). Therefore, the above equations can give a rough approximation of the minimum time wasted from the beamforming process of IEEE 802.11ad.

\subsection{Limitations of IEEE 802.11ad with vehicular networks}
As described in the standard, the BI length is limited to \SI{1000}{\milli\second}~\cite{standard}. With respect to the surrounding environment, the length can be optimised to achieve the best performance. Longer intervals increase throughput reducing the management frame transmission, however the system becomes intolerant to the delay spread. Misalignments between the TX and RX antennas can lead to more severe delay spread and consequently degradation in the performance. A typical BI length for indoor environments (zero or low mobility) is around \SI{100}{\milli\second}. However, for moving vehicles  more frequent beam switching is required ($<$\SI{30}{\milli\second}).

\begin{figure}[t]     
\centering
\includegraphics[width=1\columnwidth]{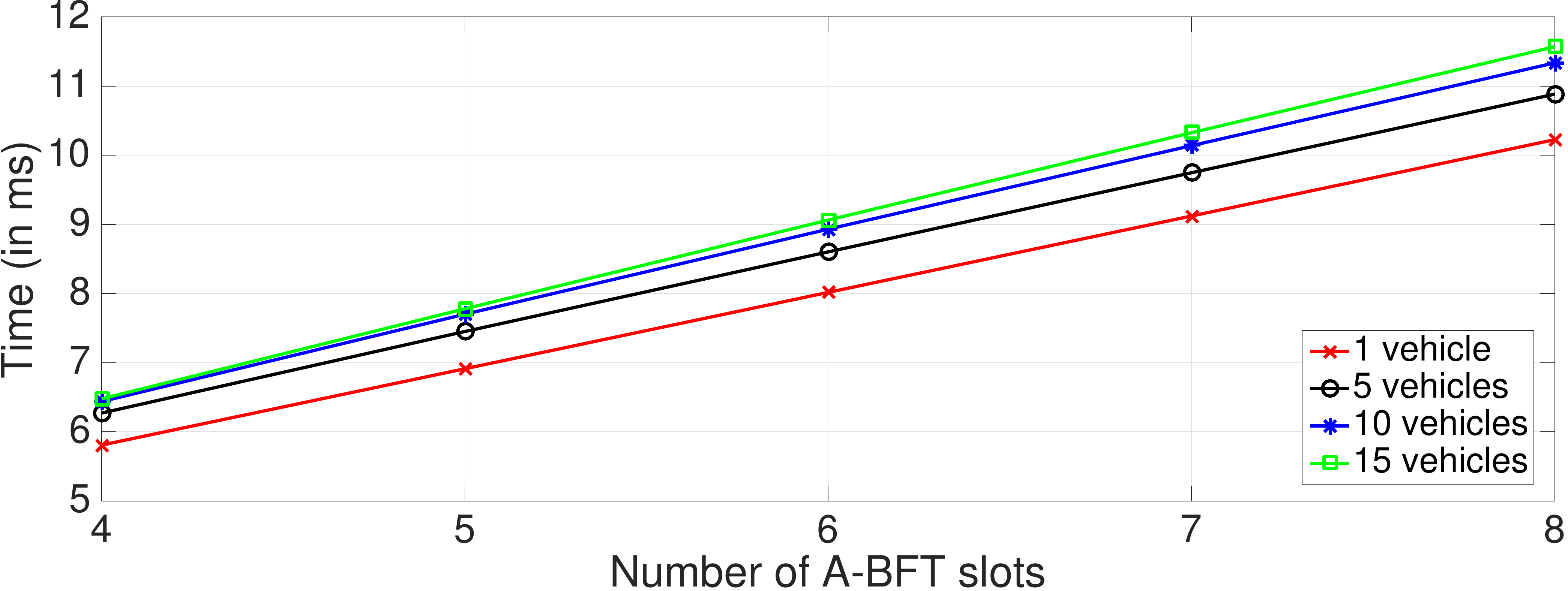}
    \caption{Example of the average delay introduced every BI from legacy beamforming training. 16 "virtual" antenna sectors were used for both TX and RX antennas and a different number of vehicles was considered.}
    \label{fig:misspent}
\end{figure}

As described before, the bidirectional exchange of training frames increases the overhead delay. Shorter BIs will lead to a bigger portion of the interval being waisted for beamforming. Additionally, increasing the number of vehicles within the coverage region of a Road Side Unit (RSU) will lead to more collisions and consequently less trained antennas. An example of the overhead delay can be seen in Fig.~\ref{fig:misspent}. More than one-third of the BI is misspent for beamforming under a vehicular scenario reducing the system throughput. 

The strict Quality-of-Service (QoS) requirements for the next-generation automotive applications require \emph{tactile-like} end-to-end delays ($<$\SI{10}{\milli\second}) and as shown (Fig.~\ref{fig:misspent}), IEEE 802.11ad cannot compensate with that. To that extent, we introduce a new beamforming approach not relying on the in-band information exchanged. Achieving overhead-free beamforming, it will be able to improve the performance and prove the capacity required for a viable solution for vehicular communications.

\section{HetNet DSRC/mmWave Beamforming for ITSs}\label{sec:het_net}

\begin{figure}[t]     
\centering
    \includegraphics[width=0.9\columnwidth]{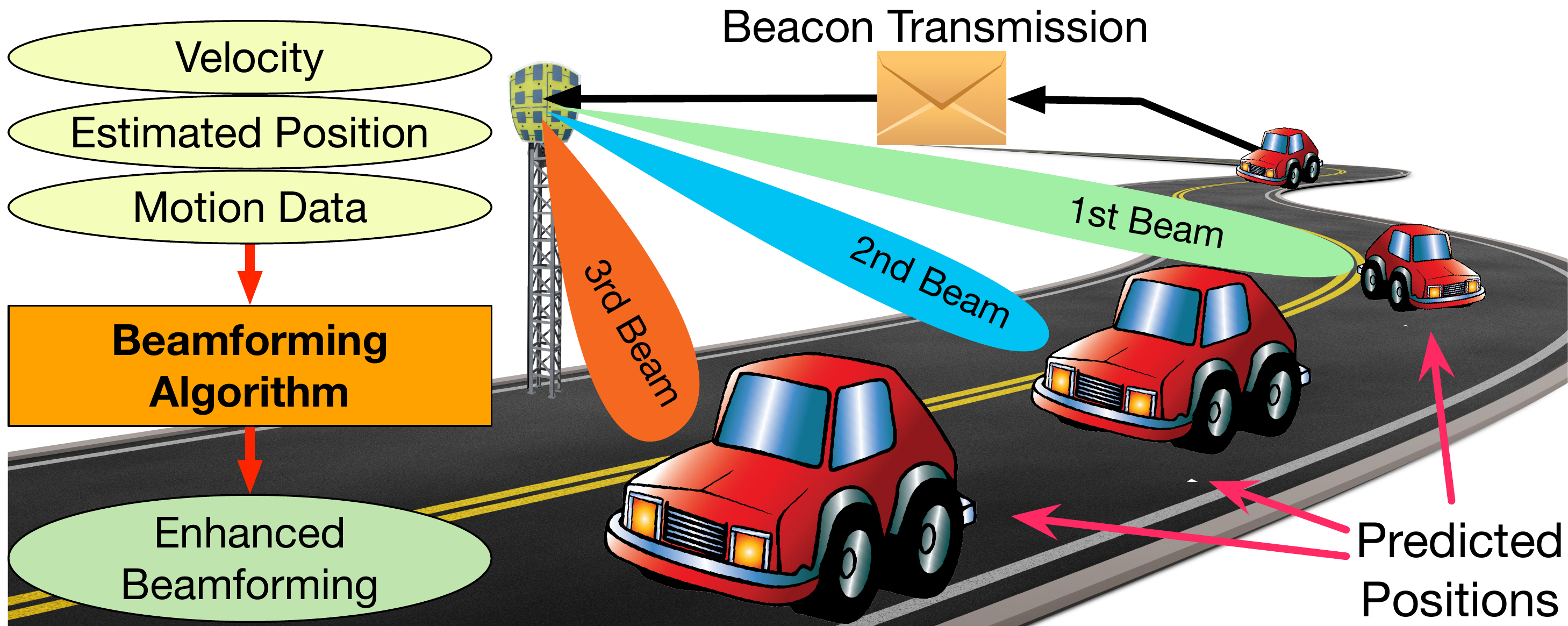}
    \caption{Smart Beamforming System: Position and motion information encapsulated in beacons are used for enhanced beamforming. RSU beams track the vehicle predicting its estimated position.}
    \label{fig:beam_alignment}
\end{figure}

Leveraging from the position and the motion information broadcast with DSRC beacons, our algorithm can provide an overhead-free beamforming. With zero-overhead, the association delays can be minimised, the beam misalignments are reduced and the performance of a mmWaves network can be enhanced. The algorithm operates as shown in Fig.~\ref{fig:beam_alignment}.

Briefly, when a vehicle approaches a RSU, prepares a bundle of information consisting of its \emph{estimated position}, its \emph{motion data} (based on the \emph{vehicle motion dynamics}) and its \emph{velocity}. These information encapsulated in a DSRC beacon are  broadcast to the nearest RSUs. The information are updated periodically and broadcast every \SI{100}{\milli\second} (DSRC beacon interval~\cite{packet_delivery}).

On the RSU side, when the initial beacon is received, the RSU aligns its beam with the vehicle according to the estimated position. When a new beacon arrives, it is examined whether the position of the vehicle has changed. If so, the beams are realigned appropriately based on the new position. Otherwise (if the beacon is lost or the position is not updated), the new position is predicted based on the received motion data.

\subsection{Position error and Mobility Model}\label{sub:pos_pred}

The position of a vehicle is estimated with respect its GPS coordinates. GPS receivers introduce a position error. As shown in~\cite{gps_accuracy}, the mean error value is \textasciitilde\SI{3}{\meter} and the standard deviation is \textasciitilde\SI{1}{\meter}. These values were measured for very large regions consisting of different kind of environments. Subject to urban scenarios specifically, the effect of an urban canyon appears due to the height of the building and GPS errors tend to become worse. Recent studies though, showed that combining the motion data with the position information with means of data fusion algorithms, can significantly reduce the position error. An example can be found in~\cite{cent_accuracy}, where authors presented an enhanced positioning system able to achieve a few-centimetre accuracy under urban scenarios, proving this positioning system as a viable solution for future autonomous vehicles.

A mobility model that can accurately characterise an urban scenario with average road density, is the synchronised flow traffic model~\cite{sync_flow}. Vehicles follow a continues traffic flow and their velocity is averaged over a mean value $\vec{v_{avg}}$ following a Normal distribution $\vec{v}\sim \mathcal{N}(\vec{v_{avg}},2)$. No significant stoppages occur and the vehicles, being limited within the road boundaries, tend to synchronise their movement performing a random manoeuvres  (change lanes, brake/accelerate smoothly).

\subsection{Beam Alignment Model based on Motion Prediction}\label{sub:align_model}

\begin{figure}[!tbp]     
\centering
	\subfloat[GPS error components and the coverage and outage intervals.]{
    \includegraphics[width=0.47\columnwidth]{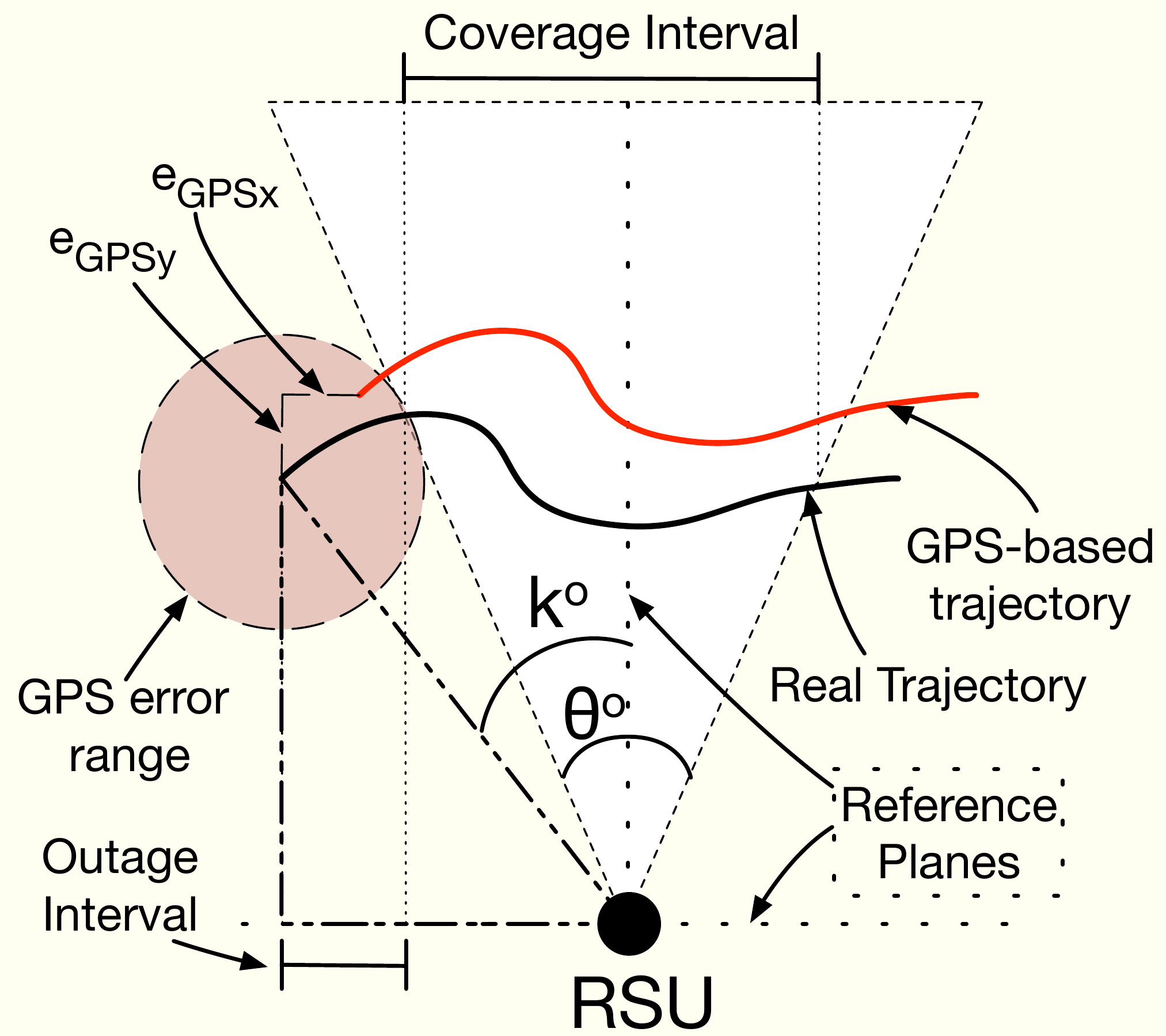}
    \label{fig:gps_error}}
    \hfill
  \subfloat[Motion of a Vehicle on the perimeter of a circle.]{\includegraphics[width=0.47\columnwidth]{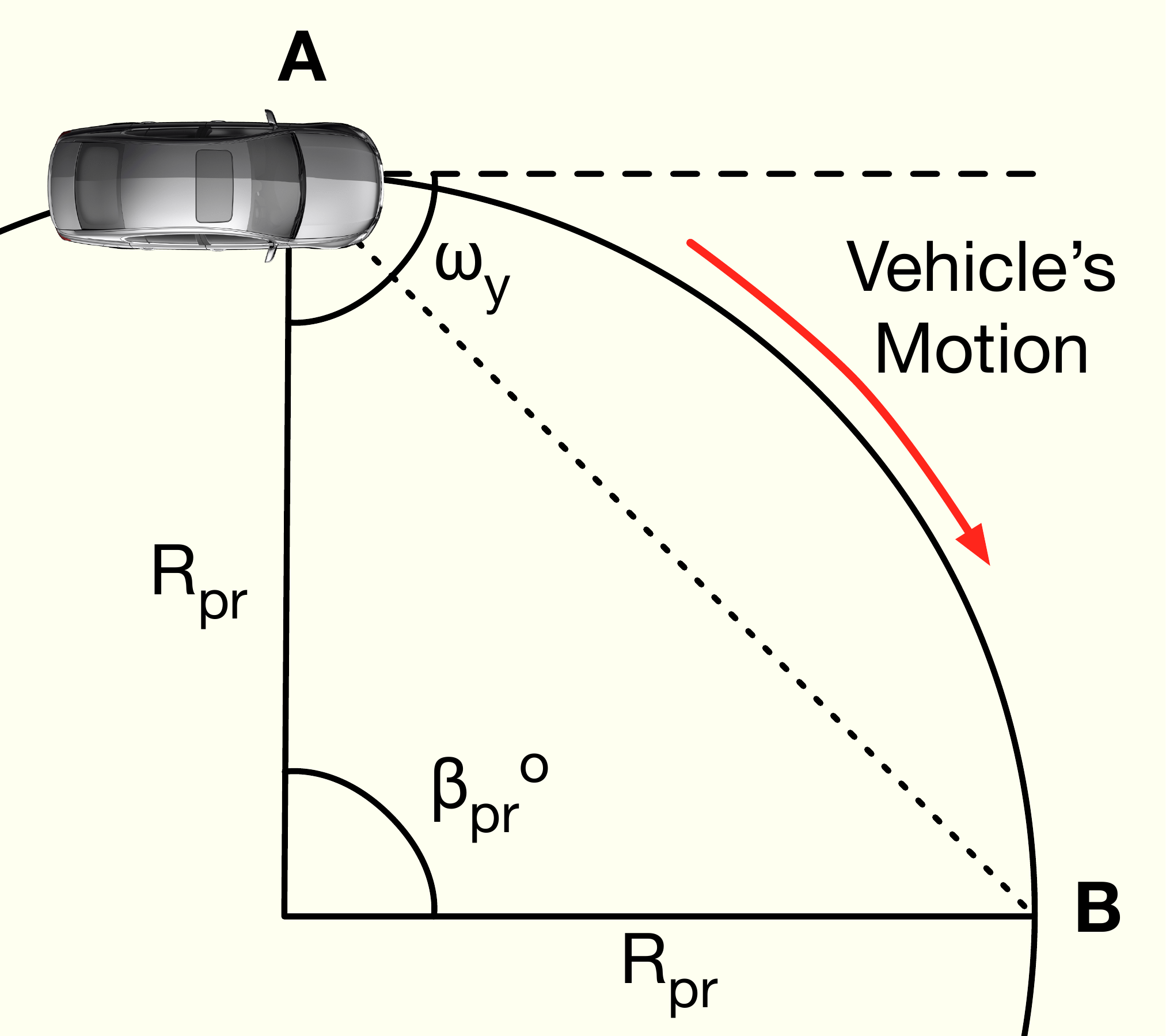}
    \label{fig:veh_motion}}
    
        \caption{a) GPS error analysis and how it influences the outage interval. Beam is steered based on the GPS trajectory. b) Motion of vehicle within a time interval used for predicting the new position.}
	\label{fig:motion_update}
\end{figure}

The estimated position is affected by the additive GPS error, i.e. $E_{pos} = C_{pos} + e_{GPS}$, where $C_{pos}$ is the real position of the vehicle and $e_{GPS}\sim\log\mathcal{N}(\mu,\sigma^2_v)$ is the log-Normal error. $\mu$ and $\sigma_v$ are the non-logarithmetised values for the mean and the variance of the log-Normal distribution. When the beam alignment is based only on the estimated position (e.g. initial beacon received), the angle $k\degree$ with respect to the reference plane is calculated (using the trigonometric equations for the right-angled triangles) and then the beam is steered accordingly (Fig.~\ref{fig:gps_error}).

For the case that no new beacon is received or the position data are outdated, the motion of the vehicle is predicted based on the motion sensory data. CAVs can be equipped with numerous sensors such as magnetometers, gyroscopes, and accelerometers. Their sensory data can be combined with data fusion algorithms, representing the angular speed of a vehicle. The angular speed is decomposed into three distinct axis, i.e. \emph{yaw $\omega_y$ for the vertical-axis, pitch $\omega_p$ for the transverse-axis, roll $\omega_r$ for the longitudinal-axis}. The above represent the rate of the angular displacement around these axis per unit time (measured in \SI{}{\deg\degree\per\second}).

Consider a constant angular velocity. A vehicle in the real world follows the surface of a sphere within one time interval. However, a vehicle moving on a road within a short period of time can only significantly change its direction  on that road plane. Therefore, the road and the vehicles will be considered as 2D objects and the movement of the vehicle will occur only on the azimuthal plane. For constant velocity a vehicle will follow the perimeter of a circle (Fig.~\ref{fig:veh_motion}). The distance between $A$ and $B$ is the distance travelled within a unit of time. This distance, as well as the angle $\beta_{pr}\degree$ can be calculated as follows:
\begin{equation}\label{eq:eq5}
\beta_{pr}\degree = m\widearc{AB} = 2 \cdot \omega_y \cdot t_{pr}
\qquad
l\widearc{AB}_{pr} = \vec{v} \cdot t_{pr}
\end{equation}
where $t_{pr}$ is the time elapsed from the latest received beacon. Using the circle properties and the outcome of equation \ref{eq:eq5}, the radius of the circle and the chord length between $A$ and $B$ can be defined as follows:
\begin{equation}\label{eq:eq6}
R_{pr} = \dfrac{180\degree \cdot l\widearc{AB}_{pr}}{2 \pi \omega_y t_{pr}}
\qquad
\overline{AB}_{pr} = 2 \cdot R_{pr} \cdot \sin(\omega_y \cdot t_{pr})
\end{equation}
where $R_{pr}$ is the radius of the circle and $\overline{AB}_{pr}$ is the distance between points $A$ and $B$. Finally, the predicted position $P_{pos}$ is given as:
\begin{equation}\label{eq:eq7}
P_{pos}(x,y) = \begin{cases}
	P_{pos}(x) = E_{pos}(x) + \overline{AB}_{pr} \cdot \sin(\beta_{pr}\degree) \\
	P_{pos}(y) = E_{pos}(y) + \overline{AB}_{pr} \cdot \cos(\beta_{pr}\degree)
	\end{cases}
\end{equation}

The model can be easily transformed to a 3D scenario, by modifying equations~\ref{eq:eq5},~\ref{eq:eq6} and~\ref{eq:eq7} to fit a spherical object.

\begin{figure*}[t]     
\centering
\includegraphics[width=0.9\textwidth]{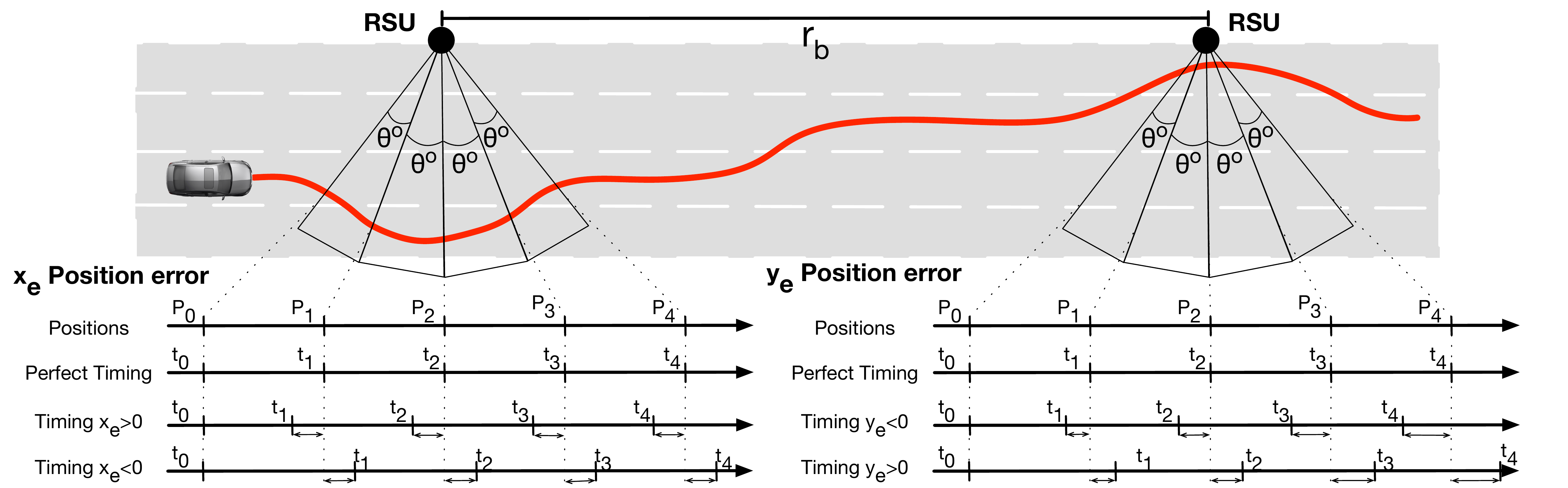}
    \caption{Vehicle performs a random movement on the road. Beam misalignments occur due the position error and the individual error components affect the misalignment differently. The movement of the vehicle is projected on a straight line.}
    \label{fig:scenario}
\end{figure*}

\subsection{Relation between position error, beamwidth and velocity}\label{sub:relationship}
Consider a scenario where a number of RSUs are placed at the side of the road and one vehicle performing a movement as in Fig.~\ref{fig:scenario}. In an ideal scenario with zero position error, the beamforming algorithm presented can achieve the maximum system performance with no outages (perfect prediction of the motion). However, GPS devices are imperfect and will lead to \emph{outage interval}. The total time that a vehicle travels within the beam (\emph{coverage interval}) is proportion to the beamwidth and the velocity and can be expressed as a function of them $T(\vec{v},\theta\degree)$ (Fig.~\ref{fig:gps_error}). 

Distinct errors will affect the system differently. For any given position error, there will be an interval that there is outage. The length of that interval is related with the position error and is minimised when it is zero. However, zero error cannot be achieved in real-world systems. The GPS error, as shown in~\cite{gps_accuracy}, can be decomposed into $(x_e,y_e)$ error components (easting and northing). Each individual errors, with respect to the RSU position, will influence the system in a different way. For example, in Fig.~\ref{fig:gps_error}, the $x_e$ will cause a more significant outage than the $y_e$.

For non-consistent position errors, an analysis of the influential error components is required. Based on this analysis, severe performance degradations can be prevented either utilising different network topologies or developing error correction algorithms. For example, for a given non-zero error causing misalignments, there is always an optimal non-zero beamwidth with regard to this error that maximises the system performance~\cite{beam_opt}.

\subsection{Antenna Gain and Beamwidth Relationship}\label{sub:antenna}
To maximise the performance with respect to the beamwidth, an antenna model forming a relationship between the gain and beamwidth should be derived a first. In this work, an ideal beam is assumed with uniform gain and no sidelobes. This model can be easily replaces with a better approximation for a specific type of antennas, to achieve more accurate results. The directivity of an antenna, associated with the beam solid angle $\Omega_A$ is given as follows~\cite{balanis}:
\begin{equation}
D = \frac{4\pi}{\Omega_A}
\end{equation}
For this ideal beam representation, $\Omega_A$ can be approximated as $\Omega_A\approx\theta_{1r}\theta_{2r}$ where $\theta_{1r}$ and $\theta_{2r}$ are the half-power (\SI{-3}{\dB}) beamwidths (in radians) of the elevation and azimuthal polarisation planes respectively.

The antenna gain $G$ can be given as~\cite{balanis}:
\begin{equation}
G = \eta D
\end{equation}
where $\eta$ is the efficiency of the antenna associated with the antenna aperture. For an ideal antenna, the efficiency is equal to $100\%$ and the gain becomes equal to the directivity. What is more, for an ideal beam $\theta_{1r}=\theta_{2r}$. So, from the above, the equation for the antenna gain with respect to the beamwidth $\theta$ (measured in degrees) becomes as:
\begin{equation}
G = \frac{4 \cdot 180^2}{\theta^2 \pi}
\end{equation} 

\subsection{Signal-to-Noise Ratio and Link Budget Analysis}

The received \emph{SNR} for the above antenna can be expressed as the ratio of the \emph{received power} $P_{rx}$ over the \emph{noise power} $P_{noise}$. The \emph{received power} is given as follows~\cite{prediction_model}:
\begin{equation}
P_{rx} = P_{tx} + G_{tx} + G_{rx} - PL
\end{equation}
where $P_{tx}$ is the transmitted power and $G_{rx}$ and $G_{tx}$ are the antenna gains for the receiver and the transmitter respectively. For this model, both antennas have ideal beams and the same beamwidth, therefore both antenna gains are equal $G_{tx}=G_{rx}$. Finally, $PL$ is the \emph{path-loss component} and is defined as:
\begin{equation}
PL = 10 \cdot n \cdot \log_{10}d + SF + C_{att} + A_{att} + R_{att}
\end{equation}
where $n$ is the path-loss exponent and $d$ is the distance separation between the RSU and the vehicle. $SF$ is the random shadowing effect following a log-Normal distribution $SF\sim\log\mathcal{N} (0,\sigma^2_{SF})$, with $\sigma$ equal to 5.8~\cite{sigma_sf}. $A_{att}$ and $R_{att}$ are the average atmospheric and rain attenuation, respectively. Finally, $C_{att}$ is a constant, representing the channel attenuation for a LOS link in an urban environment, measured at \SI{20}{\meter}~\cite{prediction_model}. For this model, a LOS link is always assumed.

The $P_{noise}$ can be calculated as:
\begin{equation}
P_{noise} = N_{floor}+10\log_{10}B+NF
\end{equation}
where $N_{floor}$ is the noise floor value, $NF$ is the noise figure, and $B$ is the antenna bandwidth. The antenna gain is associated with the beamwidth as described in Sec.~\ref{sub:antenna}. The distance between the RSU and the vehicle changes over time. For a given time $t$, the estimated position of the vehicle is known and the distance can be easily calculated. Therefore, the SNR can be expressed as a function of the beamwidth and the time as follows:
\begin{equation}
SNR(t,\theta\degree) = \frac{P_{rx}(t,\theta\degree)}{P_{noise}}
\end{equation}

The rest of the variables are always considered as constant for this model.

\subsection{Sensitivity analysis for individual error components}\label{sub:sensitivity}
The instantaneous channel capacity for a given beamwidth and time can be calculated from the Shannon-Hartley theorem as follows:
\begin{equation}
C(t,\theta\degree) = B \cdot \log_2(1+SNR(t,\theta\degree))
\end{equation}
With regard to the prediction model introduced, the beam is steered when the vehicles reaches the edge of the beam. The interval between two beam realignments is the time between $t_i$ (the system switches to the $i^{th}$ beam) and $t_{i+1}$ (the beam is realigned). For each $t_i$, a position $P_i$ exists being the real position of the vehicle (Fig.~\ref{fig:scenario}). 

Consider the same scenario as before (Sec.~\ref{sub:relationship}). Even though the vehicle performs a curved movement (Fig.~\ref{fig:scenario}), in the long-term tends to fend off the starting point, oscillating with respect to the $x$-axis. Given that only the position error exists (other sensors feedback ideal values), the real movement of the vehicle will be identical with the one predicted but shifted on the two axis. To that extent, the movement within the time interval $[t_i,t_{i+1}]$ is assumed to be a straight line. In contemplation to that, the data rate for a given $\theta\degree$ and a time interval $[t_i,t_{i+1}]$ can be calculated as:
\begin{equation}\label{eq:data_rate}
D_i(t,\theta\degree) = \int_{t_i}^{t_{i+1}} B \cdot \log_2(1+SNR(t)) dt
\end{equation}

However, due to the position error, the beamforming timing will be imperfect leading to outages. The two error components be calculated as follows:
\begin{equation}
P_e(x_e,y_e) = \begin{cases}
	x_e = x_{est} - x_r \\
	y_e = y_{est} - y_r
	\end{cases}
\end{equation}
where ($x_r$,$y_r$) is the real position of the vehicle and ($x_{est}$,$y_{est}$) is the acquired estimated position. Both errors can be divided in two cases: 1) when $x_e\geq0$ and $x_e<0$ and 2) when $y_e\geq0$ and $y_e<0$. For example, if $x_e>0 \Rightarrow x_{est}>x_r$, meaning that the beam steering will be delayed creating an outage. Therefore, the moment of the $i^{th}$ beam alignment is given as:
\begin{equation}
\widehat{t_i} = \dfrac{\sqrt[]{((P_{ix}-P_{0x})+x_e)^2 + ((P_{iy}-P_{0y})+y_e)^2}}{\vec{v}}
\end{equation}
and so, the data rate (equation~\ref{eq:data_rate}) should be calculated for the portion on the interval that the beam is aligned $[\widehat{t_i}, \widehat{t_{i+1}}]$. The above are valid only when there is at least a very short interval where there is no outage. In the case of total misalignment, the data rate is equal to zero.

Assuming that the error components are not so severe to cause total misalignment, their impact on the system performance can be analysed. \emph{Differential sensitivity analysis}~\cite{sens_analysis} was used, meaning that the sensitivity coefficient $U$ for a particular independent variable is calculated from the partial derivative of the dependent variable with respect to the independent variable. For a predefined beamwidth and denoting one of the errors as constant, the relationship of the individual uncertainty component with the channel capacity for the $i^{th}$ beam can be calculated as follows:
\begin{equation}\label{x_error_partial}
U_i(e^*) = \frac{\partial D_i(t|e^*,\theta\degree)}{\partial e} = \frac{\partial}{\partial e} \int_{\widehat{t_i}}^{\widehat{t_{i+1}}} B \cdot \log_2(1+SNR(t)) dt
\end{equation}

The formula of the channel capacity within the integral in equation~\ref{eq:data_rate} has an anti-derivative. To that extend, and denoting it as $c(t)$, can be calculated using the fundamental theorem of calculus:
\begin{equation}
D_i(t|e^*) = \int_{\widehat{t_i}}^{\widehat{t_{i+1}}} c(t) dt = C(\widehat{t_{i+1}}) - C(\widehat{t_i})
\end{equation}
Now, using the chain rule for the partial derivative (equation~\ref{x_error_partial}) we have:
\begin{equation}
U_i(e^*) = \frac{\partial D_i(t|e^*)}{\partial e} = C'(\widehat{t_{i+1}})\frac{\partial (\widehat{t_{i+1}})}{\partial e} - C(\widehat{t_i})\frac{\partial (\widehat{t_i})}{\partial e}
\end{equation}

The above equations can be numerically evaluated calculating the sensitivity coefficient for both the error components $x_e$ and $y_e$. Analysing the individual errors before developing a new ITS can significantly enhance the system performance. The impact of systematic errors can be confined to maximise the performance. This can be done by many ways. For example, changing the position of the infrastructure RSU devices, a system will easier compensate with these errors. Another solution is the development of correction algorithms able to tackle specific errors. For this work, a 2D representation was used for our system. GPS northing and easting error will not have significant differences in a real-world system. However, the same analysis can be applied to a 3D system as well, where the errors between the azimuth and the elevation plane have significant differences~\cite{gps_accuracy}. 

\subsection{Beamwidth Optimisation}
Nevertheless, reducing the influence of the position error will increase the system performance. However, any existing error will always lead to an outage. To increase the performance even more, the optimum beamwidth will be calculated for an a priory known error. $P_e$ is a random variable, so for the long-term average value, the maximum data rate $\argmaxl_{\theta\degree} D_i(t,\theta\degree|P_e)$ can be given as the expectation of the data rate denoting that it is averaging over $P_e$:
\begin{equation}\label{eq:argmax}
\hat{\theta\degree} = \argmaxl_{\theta\degree} \mathbb{E}_{P_{e}}[D_i(t,\theta\degree)]
\end{equation}

$P_e$ is decomposed in $x_e$ and $y_e$ which are both continuous random variables. Therefore, $\mathbb{E}_{P_{e}}[\cdot]$ is a positive linear function and the equation~\ref{eq:argmax} can be rewritten as:
\begin{equation}
\hat{\theta\degree} = \argmaxl_{\theta\degree} \mathbb{E}_{x_{e}}[D_i(t,\theta\degree)] + \argmaxl_{\theta\degree}\mathbb{E}_{y_{e}}[D_i(t,\theta\degree)]
\end{equation}

Denoting $f(x_e)$ and $f(y_e)$ as the probability distribution functions for $x_e$ and $y_e$ respectively, the two expected values can be calculated as:
\begin{equation}
\mathbb{E}_{x_{e}}[D_i(t,\theta\degree)] = \int_{-\infty}^{\infty} D_i(t,\theta\degree)f(x_e) dx_e
\end{equation}
\begin{equation}
\mathbb{E}_{y_{e}}[D_i(t,\theta\degree)] = \int_{-\infty}^{\infty} D_i(t,\theta\degree)f(y_e) dy_e
\end{equation}

However, the above equations should be limited to consider only the interval that there is no outage. Therefore, to properly calculate the expected value of the data rate over the $P_e$, the limits should be updated accordingly. 

Total misalignment happens when the magnitude of the $P_e$ is greater than the distance from one edge of the beam to the other. In the time domain this can be expressed as $\widehat{t_i}>t_{i+1} \Leftrightarrow \norm{\widehat{P_i}}/\vec{v} > \norm{P_{i+1}}/\vec{v}$ for $x_e<0$ and $y_e\geq0$ and $t_i>\widehat{t_{i+1}} \Leftrightarrow \norm{P_i}/\vec{v} > \norm{\widehat{P_{i+1}}}/\vec{v}$ for $x_e\geq0$ and $y_e<0$. From the above it can be calculated that the total misalignment conditions are:
\begin{equation}
  x_e = \begin{cases}
     x_e > P_i+P_{i+1},~~~~\text{for}~x_e\geq0\\
     x_e < P_{i+1}-P_i,~~~~\text{for}~x_e<0
  \end{cases}
\end{equation}
\begin{equation}
  y_e = \begin{cases}
     y_e > P_i+P_{i+1},~~~~\text{for}~y_e<0\\
     y_e < P_{i+1}-P_i,~~~~\text{for}~y_e\geq0
  \end{cases}
\end{equation}
So, using the above limits we have:
\begin{equation}
	\begin{split}
	\mathbb{E}_{x_{e}}[D_i(t,\theta\degree)] = \int_{0}^{P_i+P_{i+1}} D_i(t,\theta\degree|x_e\geq0)f(x_e) dx_e \\
	+ \int_{P_{i+1}-P_i}^{0} D_i(t,\theta\degree|x_e<0)f(x_e) dx_e
	\end{split}
\end{equation}
\begin{equation}
	\begin{split}
	\mathbb{E}_{y_{e}}[D_i(t,\theta\degree)] = \int_{0}^{P_{i+1}-P_i} D_i(t,\theta\degree|y_e\geq0)f(y_e) dy_e \\
    + \int_{P_i+P_{i+1}}^{0} D_i(t,\theta\degree|y_e<0)f(y_e) dy_e
    \end{split}
\end{equation}

The above equations can be numerically evaluated and the optimum beamwidth can be found for a given position error.

\section{Simulations and Numerical Analysis}\label{sec:results}

\begin{table}[t]
\renewcommand{\arraystretch}{1.2}
\centering
    \caption{List of Simulation Parameters.}
    \begin{tabular}{rl|rl}

    \textbf{Parameter}  &           & \textbf{Value}    & \\ \hline \hline
    Carrier Frequency   & $f_c$     & \SI{60}           & \SI{}{\giga\hertz}  \\ 
    Bandwidth 		    & $B$       & \SI{2.16}         & \SI{}{\giga\hertz} \\ 
    Path-Loss Exponent  & $n$       & 2.66              &      \\ 
    Atmospheric Attenuation  & $Atm_{att}$  & \SI{15}   & \SI{}{\dB\per\kilo\meter}      \\ 
    Rain Attenuation    & $Rain_{att}$       & \SI{25}  & \SI{}{\dB\per\kilo\meter} (in the UK)     \\ 
    Channel Attenuation & $Ch_{att}$  & \SI{70}       & \SI{}{\dB}~\cite{prediction_model}  \\
    Transmission power  & $P_{tx}$  & \SI{10}           & \SI{}{\dBm}  \\ 
    Road Block Length 	& 	$r_b$	& \SI{40}           & \SI{}{\meter} \\
    Noise Figure 		& $NF$ 		& \SI{6}            & \SI{}{\dB}  \\
    Noise Floor 		& $N_{floor}$ & \SI{-174}       & \SI{}{\dBm}  \\

    BI IEEE 802.11ad & & \SI{30}                   & \SI{}{\milli\second} \\ 
    DSRC beacon interval & & \SI{100}                   & \SI{}{\milli\second} \\ 
    GPS update interval & & \SI{1000}                   & \SI{}{\milli\second} \\ 
	\end{tabular}
\label{tab:parameters}
\end{table}

Consider a scenario where a vehicle travels on a road section with four lanes and a lane width of \SI{3.5}{\meter}. The distance travelled is one \emph{road block} $r_b$ length and the vehicle moves with random motion, as described in Sec.~\ref{sub:align_model}, and constant speed.

\begin{figure}[t]     
\centering
  \includegraphics[width=1\columnwidth]{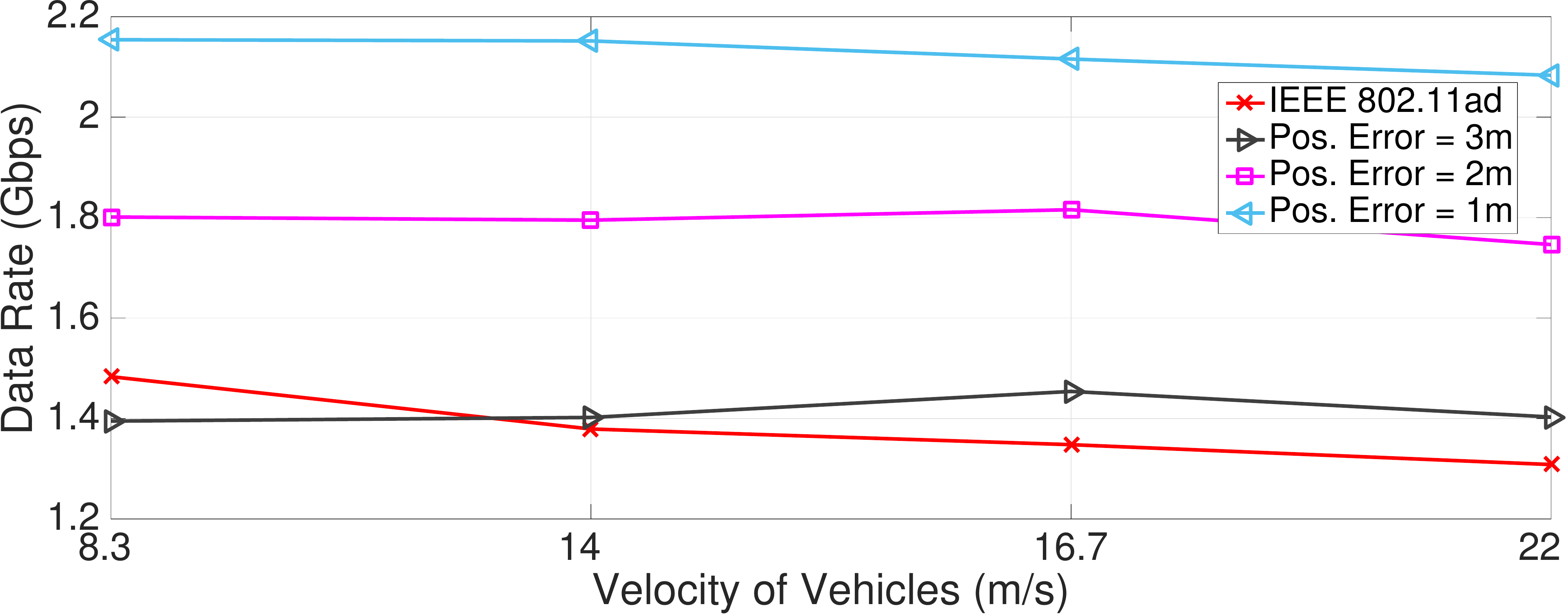}
    \caption{Average Network Throughput for different position position errors and velocities, compared to legacy IEEE 802.11ad.}
    \label{fig:diff_gps_errors}
\end{figure}

In Fig.~\ref{fig:diff_gps_errors}, the average network throughput is presented for different GPS errors and velocities. Based on the sensitivity power levels stated in IEEE 802.11ad~\cite{standard}, we used different Modulation and Coding Schemes (MCSs) with respect to the SNR. The system performance for a mean error of \SI{3}{\meter} is comparable with IEEE 802.11ad. However, when the error is reduced our algorithm significantly outperforms the legacy beamforming technique.

\begin{figure}[t]     
\centering
  \includegraphics[width=1\columnwidth]{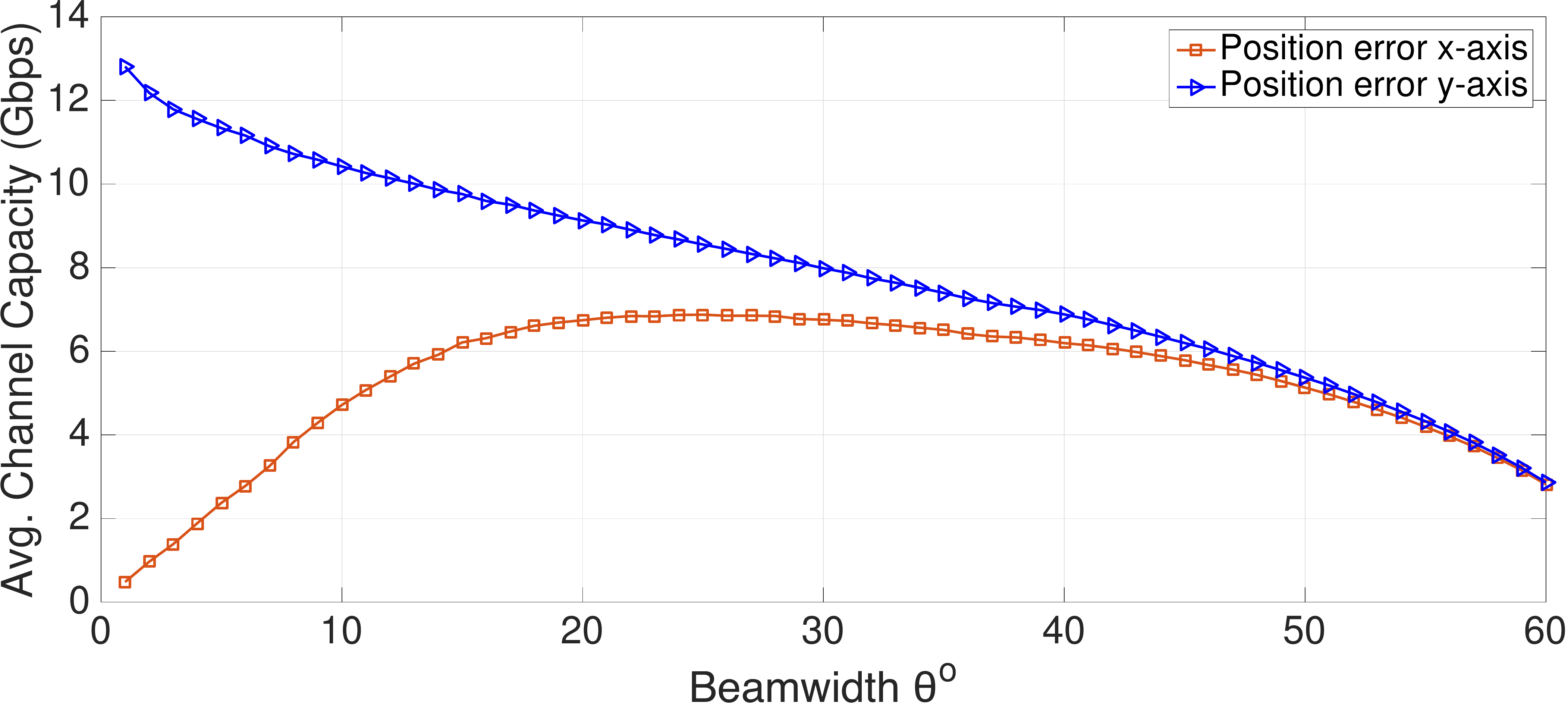}
    \caption{Differential Sensitivity Analysis of the individual position errors.}
    \label{fig:sensitivity}
\end{figure}

Using the equations in Sec.~\ref{sub:sensitivity}, the influence of each individual error component was evaluated. For each error taking random values, the other one is considered as constant. With respect to  Sec.~\ref{sub:sensitivity}, each error can take either positive or negative values and as shown in Fig.~\ref{fig:gps_error} and it has a \emph{mean absolute distance} from the real position. Therefore, the constant error consider equal to zero for each case. During this scenario, a vehicle travelling with constant speed was considered (\SI{14}{\meter\per\second}) on the same road section as before. As it can be seen from Fig.~\ref{fig:sensitivity}, even though both errors have the same magnitude, the error on the $x$-axis influences more the system performance than the one on the $y$-axis. Analysing an a priory known position error for a specific road and knowing the affect of each component is very important for the initial planning of an ITS. With better design, will be able compensate with different kind of errors, achieving better performance and enhancing the road safety.

\begin{figure}[t]     
\centering
  \includegraphics[width=1\columnwidth]{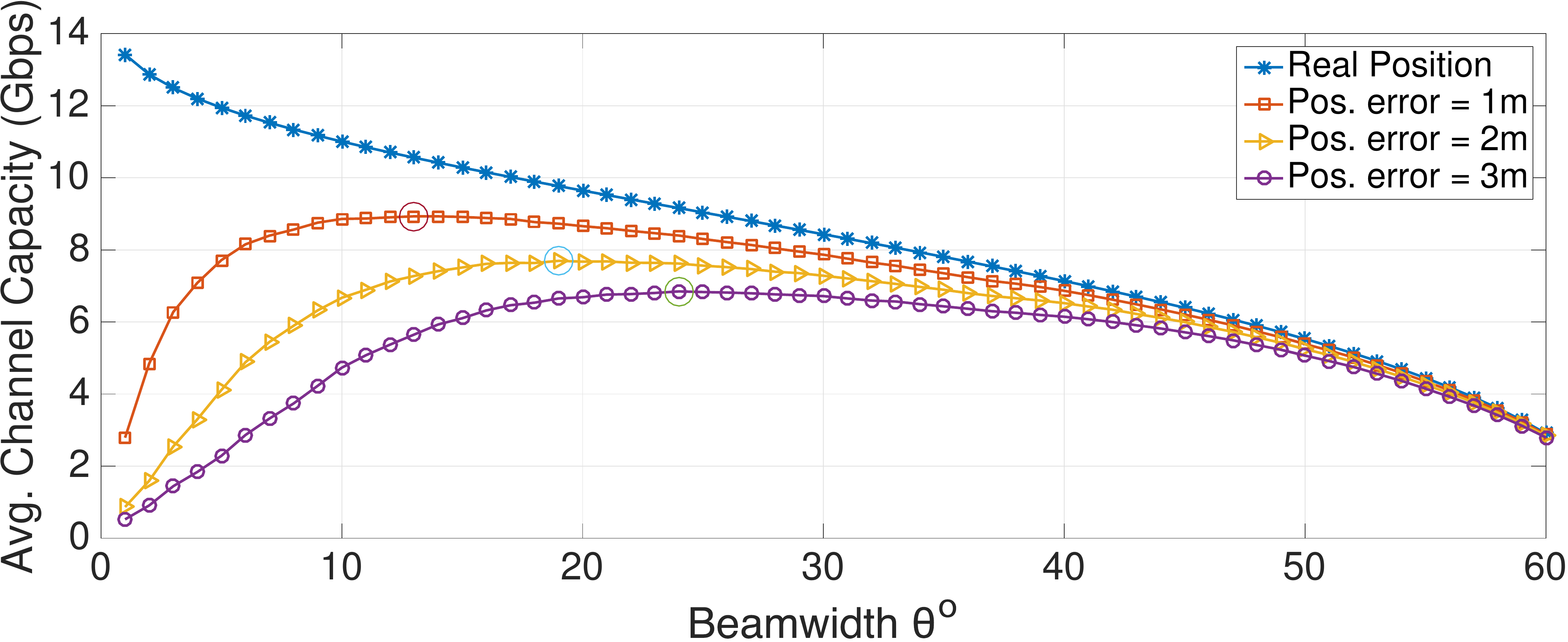}
    \caption{Optimum Beamwidth values for different position errors compared to the maximum system capacity (given when using the real position).}
    \label{fig:opt_beamwidth}
\end{figure}

Finally, Fig.~\ref{fig:opt_beamwidth} presents the average channel capacity and the optimum beamwidth for each error. The results are compared with the ideal case, where the estimated position matches the real one, i.e. zero-error exists. A mean error of \SI{3}{\meter} was considered for this scenario. As shown, when $\theta\degree$ tends to zero, even the smallest error can cause a total misalignment - thus, degrading the performance. On the other hand, when $\theta\degree$ is large, the SNR is decreased leading to lower channel capacity. An optimum beamwidth exists that is different for each position error, which maximises the average channel capacity and is shown as a circle for each case. To that extent, the system can be fine-tuned for a given error to achieve the maximum performance.

\section{Conclusions}\label{sec:conclusions}
In this paper, an intelligent beamforming training mechanism was presented. The proposed strategy can achieve overhead-free beamforming by exploiting the out-of-band information of a CAV, broadcast with DSRC beacons. Based on the position and the motion information, an agile motion-prediction model capable of estimating the position of the CAVs and predicting their movements was presented. 

What is more, the position error and how it affects the system behaviour was analysed. It was shown that distinct error components can differently affect the performance and they should be taken into consideration when planning an ITS. For example, the system can compensate with these errors either physically, by changing the position of the network devices, or with means of error correction algorithms that tackle these individual errors. An error compensation algorithm was presented later. As it was described, when position error exists, there is always an optimum beamwidth between the extreme values that maximises the system performance. 

Results showed that our algorithm can outperform the legacy beamforming technique of IEEE 802.11ad. To that extent, this algorithm can be a viable solution for the beamforming training of the mmWave antennas of the future ITSs. In the future the blockage effect, Vehicle-to-Vehicle (V2V) communications and a 3D representation of the system will be examined. 

\bibliographystyle{IEEEtran}
\bibliography{bib.bib,IEEEabrv}

\begin{thebibliography}{10}
\providecommand{\url}[1]{#1}
\csname url@samestyle\endcsname
\providecommand{\newblock}{\relax}
\providecommand{\bibinfo}[2]{#2}
\providecommand{\BIBentrySTDinterwordspacing}{\spaceskip=0pt\relax}
\providecommand{\BIBentryALTinterwordstretchfactor}{4}
\providecommand{\BIBentryALTinterwordspacing}{\spaceskip=\fontdimen2\font plus
\BIBentryALTinterwordstretchfactor\fontdimen3\font minus
  \fontdimen4\font\relax}
\providecommand{\BIBforeignlanguage}[2]{{%
\expandafter\ifx\csname l@#1\endcsname\relax
\typeout{** WARNING: IEEEtran.bst: No hyphenation pattern has been}%
\typeout{** loaded for the language `#1'. Using the pattern for}%
\typeout{** the default language instead.}%
\else
\language=\csname l@#1\endcsname
\fi
#2}}
\providecommand{\BIBdecl}{\relax}
\BIBdecl

\bibitem{its_appl}
K.~N. Qureshi and A.~H. Abdullah, ``A {Survey} on {Intelligent}
  {Transportation} {Systems},'' \emph{Middle-East Journal of Scientific
  Research}, vol.~15, no.~5, pp. 629--642, 2013.

\bibitem{gigabit_second}
J.~Choi, N.~Gonzalez-Prelcic, R.~Daniels, C.~R. Bhat, and R.~W. Heath~Jr,
  ``Millimeter {Wave} {Vehicular} {Communication} to {Support} {Massive}
  {Automotive} {Sensing},'' \emph{arXiv preprint arXiv:1602.06456}, 2016.

\bibitem{qos_req}
M.~Agiwal, A.~Roy, and N.~Saxena, ``Next {Generation} {5G} {Wireless}
  {Networks}: A {Comprehensive} {Survey},'' \emph{IEEE Communications Surveys
  Tutorials}, vol.~18, no.~3, pp. 1617--1655, Sep. 2016.

\bibitem{80211ad_anal}
A.~Maltsev, I.~Bolotin, A.~Lomayev, A.~Pudeyev, and M.~Danchenko, ``User
  {Mobility} {Impact} on {Millimeter-Wave} {System} {Performance},'' in
  \emph{2016 10th European Conference on Antennas and Propagation (EuCAP)},
  Apr. 2016, pp. 1--5.

\bibitem{standard}
``{IEEE} 802.11ad - {Enhancements} for {Very} {High} {Throughput} in the 60
  {GHz} {Band},'' no.~3, Mar. 2014.

\bibitem{beamforming_delay}
A.~Natarajan, S.~K. Reynolds, M.~D. Tsai, S.~T. Nicolson, J.~H.~C. Zhan, D.~G.
  Kam, D.~Liu, Y.~L.~O. Huang, A.~Valdes-Garcia, and B.~A. Floyd, ``A
  {Fully}-{Integrated} 16-{Element} {Phased}-{Array} {Receiver} in {SiGe}
  {BiCMOS} for 60-{GHz} {Communications},'' \emph{IEEE Journal of Solid-State
  Circuits}, vol.~46, no.~5, pp. 1059--1075, May 2011.

\bibitem{exhaustive}
Z.~Weixia, G.~Chao, D.~Guanglong, W.~Zhenyu, and G.~Ying, ``A new {Codebook}
  {Design} {Scheme} for {Fast} {Beam} {Searching} in {Millimeter-Wave}
  {Communications},'' \emph{China Communications}, vol.~11, no.~6, pp. 12--22,
  Jun. 2014.

\bibitem{packet_delivery}
F.~Lv, H.~Zhu, H.~Xue, Y.~Zhu, S.~Chang, M.~Dong, and M.~Li, ``An {Empirical}
  {Study} on {Urban} {IEEE} 802.11p {Vehicle}-to-{Vehicle} {Communication},''
  in \emph{2016 13th Annual IEEE International Conference on Sensing,
  Communication, and Networking (SECON)}, Jun. 2016, pp. 1--9.

\bibitem{beam_design}
V.~Va, T.~Shimizu, G.~Bansal, and R.~W. Heath, ``Beam {Design} for {Beam}
  {Switching} {Based} {Millimeter} {Wave} {Vehicle}-to-{Infrastructure}
  {Communications},'' in \emph{Proc. of IEEE ICC 2016}, May 2016, pp. 1--6.

\bibitem{beam_opt}
V.~Va and R.~W. Heath, ``Basic {Relationship} between {Channel} {Coherence}
  {Time} and {Beamwidth} in {Vehicular} {Channels},'' in \emph{2015 IEEE 82nd
  Vehicular Technology Conference (VTC2015-Fall)}, Sep. 2015, pp. 1--5.

\bibitem{gps_accuracy}
{Hughes, William J.}, ``{Global Positioning System (GPS) Standard Positioning
  Service (SPS) Performance Analysis Report},'' FAA GPS Performance Analysis
  Report, Tech. Rep.~94, July 2016.

\bibitem{cent_accuracy}
S.~Zhao, Y.~Chen, and J.~A. Farrell, ``{High}-{Precision} {Vehicle}
  {Navigation} in {Urban} {Environments} {Using} an {MEM}'s {IMU} and
  {Single}-{Frequency} {GPS} {Receiver},'' \emph{IEEE Trans. Intell. Transp.
  Syst.}, vol.~17, no.~10, pp. 2854--2867, Oct. 2016.

\bibitem{sync_flow}
B.~Kerner, ``Synchronized {Flow} as a {New} {Traffic} {Phase} and {Related}
  {Problems} for {Traffic} {Flow} {Modelling},'' \emph{Journal of Mathematical
  and Computer Modelling}, vol.~35, no.~5, pp. 481 -- 508, 2002.

\bibitem{balanis}
C.~A. Balanis, \emph{Antenna Theory: Analysis and Design, 4th Edition}.\hskip
  1em plus 0.5em minus 0.4em\relax John Wiley \& Sons, Mar. 2016.

\bibitem{prediction_model}
A.~Yamamoto, K.~Ogawa, T.~Horimatsu, A.~Kato, and M.~Fujise, ``Path-{Loss}
  {Prediction} {Models} for {Intervehicle} {Communication} at 60 {GHz},''
  \emph{{IEEE} Trans. Veh. Technol.}, vol.~57, no.~1, pp. 65--78, Jan. 2008.

\bibitem{sigma_sf}
M.~R. Akdeniz, Y.~Liu, M.~K. Samimi, S.~Sun, S.~Rangan, T.~S. Rappaport, and
  E.~Erkip, ``Millimeter wave channel modeling and cellular capacity
  evaluation,'' \emph{IEEE J. Sel. Areas Commun.}, vol.~32, no.~6, pp.
  1164--1179, Jun. 2014.

\bibitem{sens_analysis}
D.~Hamby, ``A {Comparison} of {Sensitivity} {Analysis} {Techniques},''
  \emph{Health physics}, vol.~68, no.~2, pp. 195--204, 1995.

\end{thebibliography}
\end{document}